# The Neutron Star and Black Hole Initial Mass Function

F. X. Timmes[1,2,3], S. E. Woosley[2,3], Thomas A. Weaver[3]

[1] Laboratory for Astrophysics and Space Research
Enrico Fermi Institute, University of Chicago
Chicago, IL  60637;   fxt@burn.uchicago.edu

[2] Board of Studies in Astronomy and Astrophysics
UCO/Lick Observatory, University of California at Santa Cruz
Santa Cruz, CA  95064

[3] General Studies Division, Lawrence Livermore National Laboratory
Livermore, CA  94550

# ABSTRACT

Using recently calculated models for massive stellar evolution and supernovae coupled to a model for Galactic chemical evolution, neutron star and black hole birth functions (number of neutron stars and black holes as a function of their mass) are determined for the Milky Way Galaxy. For those stars that explode as Type II supernovae, the models give birth functions that are bimodal with peaks at 1.27 and 1.76 $M_\odot$ and average masses within those peaks of 1.28 and 1.73 $M_\odot$. For those stars that explode as Type Ib there is a narrower spread of remnant masses, the average being 1.32 $M_\odot$, and less evidence for bimodality. These values will be increased, especially in the more massive Type II supernovae, if significant accretion continues during the initial launching of the shock, and the number of heavier neutron stars could be depleted by black hole formation. The principal reason for the dichotomy in remnant masses for Type II is the difference in the presupernova structure of stars above and below 19 $M_\odot$, the mass separating stars that burn carbon convectively from those that produce less carbon and burn radiatively. The Type Ib's and the lower mass group of the Type II's compare favorably with measured neutron star masses, and in particular to the Thorsett et al. (1993) determination of the average neutron star mass in 17 systems; $1.35 \pm 0.27$ $M_\odot$. Variations in the exponent of a Salpeter initial mass function are shown not to affect the locations of the two peaks in the distribution function, but do affect their relative amplitudes. Sources of uncertainty, in particular placement of the mass cut and sensitivity to the explosion energy, are discussed, and estimates of the total number of neutron stars and black holes in the Galaxy are given. Accretion induced collapse should give a unique gravitational mass of 1.27 $M_\odot$, although this could increase if accretion onto the newly formed neutron star continues. A similar mass will typify stars in the $8 \simeq 11$ $M_\odot$ range (e.g., the Crab pulsar). The lightest neutron star produced is 1.15 $M_\odot$ for the Type II models and 1.22 $M_\odot$ for the Type Ib models. Altogether there are about $10^9$ neutron stars in our Galaxy and a comparable number of black holes.

Subject headings: Stars: evolution - Stars: mass function - Stars: neutron - Stars: statistics - Black Hole physics

# 1. INTRODUCTION

Since the original calculations of Oppenheimer & Volkoff (1939) and Oppenheimer & Snyder (1939), numerical integration of the general relativistic equations of hydrostatic equilibrium have predicted a restricted range of allowed neutron star masses. Various equations of state allow gravitational masses for neutron stars between 0.1 and 3 $M_\odot$ (Wheeler 1966; Hartle 1978; Baym & Pethick 1979; Shapiro & Teukolsky 1983; Bahcall, Lynn, & Selipsky 1990; Lamb 1991; Lattimer & Swesty 1991), but currently favored equations of state (Lattimer & Swesty 1991) give a narrower range, 0.2 to 2.5 $M_\odot$, and those with kaon condensates (Thorsson, Prakash, & Lattimer 1994; Brown & Bethe 1994; Brown, Weingartner & Wijers 1995; Brown 1995) favor an even lower upper limit, $\simeq 1.5$ $M_\odot$. Neutron star masses below this range undergo explosive decompression while heavier ones collapse into black holes (Prakash, Lattimer & Ainsworth 1988; Blinnikov et al. 1990; Colpi, Shapiro, & Teukolsky 1993).

While the range of allowable masses is set by fundamental nuclear theory, the total number of neutron stars and black holes as well as their distribution with mass is a consequence of stellar evolution. A number of uncertain factors make a straightforward determination difficult – the mixing length theory for convection which all current stellar evolution codes employ, our still developing understanding of the explosion mechanism, the possibility that the star may lose a considerable, but uncertain, part of its mass during its evolution, and finally, residual uncertainty in key nuclear reaction rates. Still, stellar and supernova modeling have advanced to a state such that a preliminary estimate of what we shall term "the neutron star initial mass function" is warranted, if only to clarify and quantify some of the above uncertainties. We call it the "initial" mass function because accurate measurements of neutron star and black hole masses occur only in binary systems where the mass of the collapsed object may sometimes increase.

After briefly reviewing some of the factors that determine the neutron star mass that results from a given stellar model, we present and discuss the implications of a recent systematic study of massive stellar evolution by Weaver & Woosley (1996) and Woosley & Weaver (1995). These calculations give ejecta whose composition, when integrated into a Galactic chemical evolution model, agrees remarkably well with what is observed in the Sun and other nearby stars (Timmes, Woosley & Weaver 1995). The iron core masses predicted by these same models can be integrated, using the same codes and parameter choices (for self-consistency) as were successfully used for calculating chemical evolution, to obtain the desired birth function for compact objects.

We have tried to address in a simple way the effects of mass loss and uncertainties in the explosion mechanism. We address the former by including in our study the stellar models of Woosley, Langer, & Weaver (1993, 1995) which explicitly followed the mass loss that might be appropriate for either single or mass exchanging binary stars. The mass of the neutron star is affected only if these stars lose not only all of their hydrogen envelopes, but end up with significantly smaller helium cores. Such a reduction in helium core mass seems necessary to produce the light curves of Type Ib supernovae (Ensman & Woosley 1988). As one might expect, a reduced helium core leads to smaller neutron stars (see also Brown & Weingartner 1994). Reduced semiconvection would also lead to smaller helium cores (Weaver & Woosley 1993) and thus smaller neutron star masses, but we have not included this effect.

We address the issue of explosion mechanism and mass cut by presenting results for three possible choices of remnant mass: (1) the iron core of the presupernova star, (2) the mass interior to the oxygen–burning shell in the presupernova star, and (3) the actual remnant mass computed when the star is exploded using a piston parameterized in such a way as to impart prescribed values of total kinetic energy to the ejecta at infinity. As emphasized by Woosley & Weaver (1995), the fallback of material that slows below escape velocity due to hydrodynamic interaction far from the nascent neutron star is critical in determining the final mass. For example, three otherwise identical 35 $M_\odot$ solar metallicity stars, each a "successful" supernova with final kinetic energies at infinity of 1.23, 1.88, and 2.22 $\times 10^{51}$ ergs, leave remnants having baryonic masses of 7.38, 3.86, and 2.03 $M_\odot$. Such sensitivity to uncertain parameters renders predictions of remnant masses particularly uncertain, especially in the more massive stars. Fortunately such stars are rare. For these very massive stars we consider the remnant masses produced by a variety of final kinetic energies.

## 2. THE PATH TO INSTABILITY

Why should one expect the typical neutron star to have a gravitational mass near 1.3 ~ 1.4 $M_\odot$? There have been a number of discussions of the physics that sets the mass of the iron core that collapses as a massive star ends its life (Hoyle & Fowler 1960; Baron & Cooperstein 1990; Cooperstein & Baron 1990; Woosley & Weaver 1992,1994). We briefly review some of the main points.

During carbon and oxygen burning, pair neutrino losses lead to a sufficient decrease in the central entropy of a massive star that the concept of Chandrasekhar mass becomes, in a general sense, meaningful. Except for stars lighter than about 11 $M_\odot$, the degeneracy is not so extreme as to lead to off–center ignition or "flashes", but even for much more massive stars, the iron cores that collapse are at least mildly degenerate.

The traditional Chandrasekhar mass (Chandrasekhar 1938) is given by

$$M_{\mathrm{Ch0}} = 5.83 \; Y_\mathrm{e}^2 \;, \tag{1}$$

which for $Y_e = 0.50$ (matter composed of an equal number of protons and neutrons) is 1.457 $M_\odot$. The typical iron core at the time it collapses has a central $Y_e$ of 0.42 which rises to 0.48 at the edge of the iron core and is roughly linear in mass between. Taking $Y_e = 0.45$ as an average, one might expect a Chandrasekhar mass of 1.18 $M_\odot$. But there are numerous corrections, some of which are large, that take into account the thermal structure of the core, in particular that its entropy is not zero, the fact that the particles responsible for the pressure have charge (Coulomb corrections), the fact that the iron core is surrounded by matter (and thus has a surface boundary pressure), as well as the usual special and general relativistic corrections (Shapiro & Teukolsky 1983) that, by themselves, reduce 5.83 $Y_e^2$ to 1.42 $M_\odot$ for $Y_e = 0.50$ and 1.15 $M_\odot$ for $Y_e = 0.45$.

The pressure decrement due to electrostatic interactions among the non-uniformly distributed ions and electrons further reduce the traditional Chandrasekhar mass by an amount given approximately by

$$M_{\rm Ch} \simeq M_{\rm Ch0} \left[ 1 \ - \ 0.0226 \ \left(\frac{Z}{6}\right)^{2/3} \right] . \tag{2}$$

For a pure carbon white dwarf this reduces the Chandrasekhar mass from 1.42 $M_\odot$ to 1.39 $M_\odot$ and, for $Y_e = 0.45$ and $Z = 26$, $M_{Ch} = 1.08$ $M_\odot$ If this was all the relevant physics, neutron stars would be very light, especially after subtracting their binding energy.

But another very important correction must beapplied since the iron core mass has finite entropy. The core must lose entropy to collapse, but its entropy is not and cannot go to zero. To a first approximation

$$M_{\rm Ch} \simeq M_{\rm Ch0} \left[ 1 \ + \ \left(\frac{\pi^2 k^2 T^2}{\epsilon_{\rm F}^2}\right) \right] , \tag{3}$$

where $\epsilon_{\rm F}$ is the Fermi energy for the relativistic and partially degenerate electrons

$$\epsilon_{\rm F} \ = \ 1.11 \ (\rho_7 \ Y_e)^{1/3} \ {\rm MeV} . \tag{4}$$

The distinction between $\epsilon_{\rm F}$ and the chemical potential, $\mu_e$, is important in deriving equation (4). Roughly one has (Baron & Cooperstein 1990)

$$\mu_e \simeq \epsilon_{\rm F} \left[ 1 - \frac{1}{3} \ \left(\frac{\pi k_B T}{\epsilon_{\rm F}}\right)^2 \right] . \tag{5}$$

The effective Chandrasekhar mass may also be expressed in terms of the electronic entropy per baryon

$$M_{\rm Ch} \approx M_{\rm Ch0} \left[ 1 \ + \ (\frac{s_e}{\pi Y_e})^2 \right] , \tag{6}$$

where $s_e$ in units of the Boltzmann constant, $k$, is (Baron & Cooperstein 1990)

$$s_e = \frac{S_e}{N_A k} = \frac{\pi^2 T Y_e}{\epsilon_F} \simeq 0.50 \; \rho_{10}^{-1/3} \left(\frac{Y_e}{0.42}\right)^{2/3} T_{\text{MeV}} \; . \tag{7}$$

At the time the iron core in a 15 $M_\odot$ star collapses, the electronic entropy ranges from 0.4 in the center to 1 at the edge of the iron core. Taking 0.7 as a rough average (and again $Y_e \approx 0.45$), one has an "effective Chandrasekhar mass" of 1.34 $M_\odot$, which is in fact close to the mass of the core in this star when it does collapse (1.32 $M_\odot$). For a 25 $M_\odot$ star, the presupernova core entropy ranges from 0.5 to 1.8, suggesting a Chandrasekhar mass of 1.79 $M_\odot$. It is the well–known characteristic of the more massive stars that they have more entropy that leads the mass of the neutron star to be positively correlated, in a general way, with the main–sequence mass of the star. The entropy loss that occurs owing to neutrino losses during carbon burning, especially for those stars that experience convective carbon core burning, is particularly important in determining this critical mass.

However, having arrived at an approximate Chandrasekhar mass, we are not nearly done. There are terms, largely negligible, for pressure other than electrons, for rotation, and for the surface boundary pressure. Much more important, however, is the manner in which the star approaches the Chandrasekhar mass during silicon burning.

If there is an active burning shell within the core, it will not collapse; contraction leads to accelerated nuclear burning and expansion. However, the iron core does not grow by radiative diffusion, but by a series of convective shell burning episodes, the last of which overshoots the (generalized) Chandrasekhar mass (usually there are just one or two such episodes). How far each silicon–burning shell extends is sensitive both to the the previous entropy history in the inner regions of the star (i.e., its entire life history, especially the location of the previous oxygen–burning shells), and to how convection is treated (e.g., semiconvection or no semiconvection). This leads to some aspects of chaos and uncertainty in the presupernova iron core mass; chaos in the sense that two stars separated only by a small mass on the main sequence can have iron core masses that differ appreciably if one of them requires an additional shell–burning episode; uncertain in the sense that mixing length convection theory during silicon burning is not very accurate.

Including realistic treatments of the different pieces of physics (i.e., using the stellar models themselves), the iron core that collapses is typically between 1.25 and 2.05 $M_\odot$ (Woosley & Weaver 1995) with $1.3 \simeq 1.6$ being typical of stars with masses between 11 and 40 $M_\odot$. These iron cores, supported by the pressure of $\gamma \simeq 4/3$ electrons, collapse due to a combination of photodisintegration and electron capture (Cooperstein & Baron 1990). But before one connects these iron core masses with an observed neutron star mass, there are additional uncertainties that arise because the iron core does not necessarily equal the mass of the baryonic remnant. Because of nucleosynthetic restrictions it is a lower bound (Weaver, Zimmerman, & Woosley 1978), but additional material may accrete during the tenth of a second or so it takes the explosion to develop as well as later when the outgoing shock interacts with the mantle and envelope of the star. This will be discussed in more detail later in the paper.

Finally, the baryonic mass of the collapsed remnant must be reduced by the neutrinos that are lost. That is, the binding energy of the neutron star must be subtracted. Transformation of the baryonic remnant masses to gravitational remnant masses is approximated here using the quadratic

$$M_{baryon} - M_{grav} = \Delta M = 0.075 \, M_{grav}^2 \quad , \tag{8}$$

where the masses are measured in solar masses (Burrows & Lattimer 1986; Lattimer & Yahil 1989). The number 0.075 is close to the centroid (within 20%) of the results obtained for a large number of reasonable nuclear equations of state (J. M. Lattimer 1995, private communication). Other functional forms or coefficients for the transformation (e.g., Cook, Shapiro, & Teukolsky 1994) slightly change the values of the gravitational mass, hence the location of the peaks in the distribution functions we obtain, but does not alter the principal conclusions.

## 3. ESTIMATES OF THE REMNANT MASSES

Weaver & Woosley (1996) have followed the presupernova evolution and Woosley & Weaver (1995) the explosive evolution of 78 Type II supernovae for a grid of stellar masses and metallicities including stars of 11, 12, 13, 15, 18, 19, 20, 22, 25, 30, 35, and 40 $M_\odot$ and metallicities Z = 0, $10^{-4}$, 0.01, 0.1, and 1 times solar. In both the presupernova and exploded models the nucleosynthesis of 200 isotopes was determined. Each star was exploded using a piston to give a specified final kinetic energy of the ejecta at infinity (typically $1.2 \times 10^{51}$ ergs). The final mass of the collapsed remnant was determined and often found not to correspond to the location of the piston (typically the edge of the iron core), but to a location farther out due to the fallback of material onto the remnant. As discussed in §2, this location is sensitive to the explosion characteristics and the presupernova star.

Masses of the iron cores in the presupernova models are shown as a function of the main-sequence mass in Figure 1 for stars whose initial metallicity is zero and in Figure 2 for stars which have a solar initial composition. These iron core masses (open triangles in Figs. 1 and 2) form a lower bound for the mass of the remnant. Ejection of the neutron-rich material interior to the iron core would cause severe nucleosynthetic problems (Weaver et al. 1978). Masses of the iron core plus the mass of the silicon shell (total mass interior to the oxygen–burning shell) in the presupernova models are shown in each figure as the open squares. The inner edge of the oxygen burning shell usually has an entropy jump which might be a natural location for the mass cut. An upper bound to the remnant masses, which is less certain than the lower bound represented by the iron core mass, is the total mass interior to the oxygen–burning shell in the presupernova models. It is possible for the remnant to extend even beyond this boundary if significant fallback occurs following the explosion.

These presupernova stars were then exploded (Woosley & Weaver 1995). The baryonic mass of the remnant, which includes the fallback, is shown in Figures 1 and 2 as the filled circles. The total kinetic energy of the ejecta at infinity is labeled alongside each of the remnant masses in units of $10^{51}$ ergs. To have all the remnant masses $\simeq 1.5$ $M_\odot$ the explosion energy in the M $\geq 25$ $M_\odot$ presupernova models must be steadily increased in order to overcome the increased binding energy of the mantle. For a constant kinetic energy at infinity, the amount of matter that falls back onto the remnant in the high–mass stars increases rapidly. Unless the explosion mechanism, for unknown reasons, provides a much larger characteristic energy in more massive stars, it appears likely that stars larger than about 30 $M_\odot$ will have dramatically reduced yields of heavy elements and leave massive remnants (M $\geq 10$ $M_\odot$) which become black holes (Timmes et al. 1995). In general, the remnant masses left by the zero–metal stars are larger than those left by the solar–metallicity stars since the zero–metal stars tend to have a more compact density structure and experience less mass loss than the solar–metallicity stars at the time of core collapse. However, this singularity at zero metallicity does not characterize more massive stars having even $10^{-4}$ times the metallicity of the Sun.

The gravitational masses of the cores of all the presupernovae (Weaver & Woosley 1996) and the remnants left by the exploded models (Woosley & Weaver 1995) were calculated with equation (8). These gravitational masses were then incorporated into the Galactic stellar–chemical evolution model of Timmes et al. (1995). The formalism in the chemical evolution model is simple and relatively standard. Each radial zone in an exponential disk begins with zero gas and accretes primordial material over a 2$\simeq$4 Gyr $e$-folding timescale. The isotopic evolution at each radial coordinate is calculated using "zone" models (as opposed to hydrodynamic models) of chemical evolution. Standard auxiliary quantities such as a Salpeter (1955) initial mass function and a Schmidt (1959) birthrate function were used. Besides the Type II supernova yields of Woosley & Weaver (1995) discussed above, the model uses abundance yields for Type Ia supernovae (Nomoto, Thielemann, & Yokoi 1984; Thielemann, Nomoto, & Yokoi 1986) and intermediate–low mass stars (Renzini & Voli 1981) are also included. The abundances at 8.5 kpc (galactocentric radius) and 4.6 Gyr ago (solar age) are in excellent agreement with the Anders & Grevesse (1989) solar abundances and major features of the observed abundance histories of all elements lighter than zinc. The derived Galactic supernova rates as a function of time are in good agreement with the estimates of van den Bergh & Tammann (1991) and van den Bergh &

McClure (1994). For self-consistency, the distribution of gravitational remnant masses is calculated with the same model and parameters.

Main–sequence stars between 8 $M_\odot$ and 11 $M_\odot$ constitute a considerable fraction of stars for any reasonable initial mass function and should be included in any calculation of neutron star distribution functions. These stars occupy a transition region bounded on the lower mass end by stars that ignite carbon degenerately and on the upper mass end by stars that ignite all the major nuclear–burning stages nondegenerately in their cores. As such they have a complex structural evolution.

Some of the stars in the $8 \simeq 11$ $M_\odot$ region may become asymptotic giant branch (AGB) stars with neon–oxygen–magnesium (NeOMg) cores and lose their envelopes, leaving behind NeOMg white dwarfs (Nomoto 1987), but this fraction is likely to be small owing to the short lifetime as an AGB star (Nomoto 1984).

Stars in this narrow mass range are characterized by a degenerate core structure following carbon burning, off-center ignition, and episodes of convectively bounded flame propagation (Woosley, Weaver & Taam 1980; Miyaji, Nomoto & Yokoi 1980; Nomoto 1982; Habets 1986; Miyaji & Nomoto 1987; Nomoto 1987; Mayle & Wilson 1988; Hashimoto, Iwamoto, & Nomoto 1993; Timmes, Woosley, & Taam 1994). For example, Woosley et al. (1980) found that neon burning ignited several tenths of a solar mass out from the center. Both neon and oxygen burned away in a region from 0.3 to 1.3 $M_\odot$, leaving a peculiar inverted compositional structure where neon lay both on top of and beneath an extensive shell of silicon and intermediate–mass elements. The remaining 0.3 $M_\odot$ of neon and oxygen at the center then burned by a series of convectively bounded burning stages ("flashes") much like what has been described for carbon–oxygen dwarfs ignited at the outer edge (Timmes et al. 1994). Eventually an iron core in hydrostatic equilibrium is formed. The cores that collapse have a baryonic mass that clusters around the zero entropy, $Y_e$=0.5 value of 1.39 $M_\odot$ (the value used throughout this paper). Although the $8 \simeq 11$ $M_\odot$ range of stellar masses is narrow, these stars affect the low–mass end of the neutron star birth function. The Crab pulsar and nebula is often associated with a progenitor star in the $8 \simeq 11$ $M_\odot$ mass range (Arnett 1975; Nomoto 1985) chiefly because such objects make neutron stars and eject helium contaminated by only traces of heavier elements.

## 4. THEORETICAL EXPECTATIONS

The number of remnants as a function of *gravitational* mass is shown in the histograms of Figure 3. The number of remnants has been normalized such that the total number in all mass bins is equal to one. A mass resolution (the width of each mass bin) of 0.014 $M_\odot$ was used. The distribution obtained using as input the lower bound (iron core) masses of the presupernova models is shown in Figure 3a as filled triangles. This function has significant peaks at 1.19 $M_\odot$ and 1.58 $M_\odot$. Below 1.40 $M_\odot$ (i.e., the lower peak in Fig. 3a) the average gravitational mass is 1.23 $M_\odot$; above 1.40 $M_\odot$ (the heavier group), the average mass is 1.54 $M_\odot$. For a Salpeter (-1.35 exponent) initial mass function, the peak at 1.18 $M_\odot$ is the largest peak. Using the presupernova iron core mass plus silicon shell mass (total mass interior to the oxygen–burning shell), one obtains the histogram marked with filled squares in Figure 3b. This function has significant peaks at 1.36 $M_\odot$ and 1.90 $M_\odot$ and the peaks, respectively, have average masses 1.38 and 1.95 $M_\odot$. We do not consider the small excess of remnant masses at the upper end of the plot shown in Figures 3a and 3b to be statistically meaningful and have grouped them with the second peak.

Perhaps the most realistic distribution is the one for the remnants left by the exploded stars, which is marked with filled circles in Figure 3c. In this calculation only the largest explosion energy models were used (see Figs. 1 and 2). The explosion energy dependence is examined below. In total, these post–explosion remnants have a average mass of 1.61 $M_\odot$ with a standard deviation of $\pm$ 0.21 $M_\odot$, but again display a bimodal distribution with significant peaks at 1.27 $M_\odot$ and 1.76 $M_\odot$ (average masses 1.28 and 1.73 $M_\odot$), although these masses could be increased if significant accretion continues during the launching of the shock (e.g., Wilson et al. 1988; Wilson & Mayle 1993; Bethe 1993). Such accretion may be required to avoid nucleosynthetic difficulties in the models of Herant et al. (1994), Janka & Müller (1995) and Burrows, Hayes, & Fryxell (1995).

If the majority of stars between 8 and 11 $M_\odot$ become AGB stars, instead of Type II supernovae that leave a neutron star as assumed in Figure 3, then the amplitude of the low–mass peak (1.19 $M_\odot$ for the iron cores, 1.36 $M_\odot$ for the iron core plus silicon shell mass, and 1.27 $M_\odot$ for the post–explosion remnants) decreases substantially. In this extreme case, a Salpeter initial mass function allows the high–mass remnant peak in Figures 3a, 3b, and 3c to become the peak with the largest amplitude. Having all stars between 8 and 11 $M_\odot$ become AGB stars also makes the spread about the low–mass peak also become slighly smaller. The amplitude and spread around the peaks for prescribed fractions of stars that become either a AGB stars or Type II supernova lie intermediate to the two cases considered here.

Note that the two post–explosion peaks are between the lower bound iron cores and the rough upper bound oxygen shell masses. The lack in Figure 3c of the excess number of heavy remnants seen in Figures 3a and 3b (the "third peak") is a consequence of both the smaller remnant masses in the high–explosion energy models and the larger $\Delta M$ corrections for larger masses. These two effects conspire, by coincidence, with the stellar initial mass function to make this excess merge with the second peak. The relative amplitudes, but not the location, of the peaks shown in the figures vary with the choice of the exponent in the initial mass function (§6).

The main reason for the bimodality of the distributions shown in Figure 3 is a qualitative difference in the presupernova structure of stars above and below $\simeq 19$ M$_\odot$. Figure 4 shows the iron core masses (filled circles) and iron core masses plus the silicon shell mass (open stars) for a fine mass grid of 55 presupernova stars (Weaver & Woosley 1996). This is a much finer grid of models than was exploded and used in the present paper, but illustrates dramatically the discontinuity in iron core mass at 19 M$_\odot$. This mass marks the transition between stars that burn carbon convectively in the core and those that burn radiatively (Weaver & Woosley 1993, 1996). Below 19 M$_\odot$ the star typically experiences convective carbon core burning and three episodes of convective carbon shell burning. Above 19 M$_\odot$, however, for the chosen value of the $^{12}$C$(\alpha,\gamma)^{16}$O reaction rate (1.7 times the Caughlan & Fowler (1988) value), the abundance of carbon is too small to provide sufficient power above the neutrino losses to drive convection in the core. These stars burn carbon and neon radiatively in the core. Stars above 19 M$_\odot$ basically skip carbon burning and jump, so far as core structure is concerned, from helium depletion to oxygen ignition. The entropy thus remains high in these cores, and the adjusted Chandrasekhar masses and iron core masses are larger (Woosley & Weaver 1994). The star-to star scatter in Figure 4 is probably real. The iron cores are not monotonic with respect to mass (or metallicity). Variations are caused by the interaction of the various convective zones during the later stages of nuclear burning (Weaver & Woosley 1993, 1996) along with the corrections discussed in §2.

Unlike elemental ratios such as [Si/Fe], which are relatively insensitive to the chemical evolution model, absolute numbers, such as the total number of neutron stars and black holes in the Galaxy, are much more dependent upon assumptions and uncertainties. The uncertainties in the total baryonic mass of the Galaxy, shape of the initial mass function, integration limits of the initial mass function, and the star formation rate (perhaps the most critical parameter) all contribute to the error. In the following estimates we will use the values in Timmes et al. (1995) for these uncertain parameters. If every post–explosion remnant in Figure 3c is a neutron star, then there are a total of $1.9 \times 10^9$ neutron stars in the Galaxy. If every remnant above 1.7 M$_\odot$ turns into a black hole, then the post–explosion peak at 1.76 M$_\odot$ would be lost to black holes and there would be $4.7 \times 10^8$ neutron stars and $1.4 \times 10^9$ stellar–mass black holes in the Galaxy. In this case, only the peak at 1.27 M$_\odot$ would survive as neutron stars. The various possibilities for remnant masses that turn into black holes are indicated schematically by the arrows in Fig 3. Other cases, such as considering only the iron core masses instead of the post–explosion remnants, follow accordingly.

The explosion energy dependence of the post-explosion neutron star birth function is examined in Figure 5. The width of the mass bins is the same as in Figure 3 and the same normalization procedure followed. Using the medium and low explosion energy models gives low–mass peaks in the same mass bins as in the high–energy models of Figure 3, but different excesses at larger remnant masses. As before, these small "third peak" at large remnant masses are probably devoid of statistical significance. If most stars between 8 and 11 $M_\odot$ become AGB stars and are distributed according to a Salpeter initial mass function, then the amplitude of the at 1.27 $M_\odot$ is decreased significantly and the peak at 1.76 $M_\odot$ has the largest relative amplitude.

## 5. COMPARISON TO THE OBSERVATIONS AND THE ROLE OF MASS LOSS

More than 700 neutron stars have been identified in the Galaxy. Most of these are solitary, rotation-powered pulsars located in or near the disk, but approximately 100 are known to be in binary systems. Of those in binary systems, about 25 are rotation–powered pulsars, about 35 are accretion–powered pulsars, and the remainder have not been observed to pulse but have been identified as neutron stars by their X-ray properties (Lewin & Joss 1983; Joss & Rappaport 1984; van den Heuvel 1991; Lamb 1991; Taylor, Manchester, & Lyne 1993). So far, only 17 of these binary neutron star systems are favorable to observations that can determine the masses of the neutron stars (Thorsett et al. 1993; Brown et al. 1995).

The upper half of Figures 6a, 6b, and 6c show *some* of the currently measured neutron star masses. The most probable mass of each star is shown by the filled circle while the uncertainty is indicated by the length of the error bar. The first seven systems shown are all believed to be members of neutron star – neutron star binary systems. Doppler velocity curves can be observed with sufficient precision in some of these systems to observe relativistic effects. The masses of PSR 1913+16 and its companion PSR 1913+16C are from Taylor & Weisberg (1989). Mass estimates of PSR 1534+12 and its companion 1534+12C are from Wolszczan (1991) and Arzoumanian (1995). The mass estimate for PSR 1855+09 is from Ryba & Taylor (1991) and Kaspi, Taylor, & Ryba (1994). Masses of 2127+11 and its companion 2127+11C are from the Brown et al. (1995) compilation. The last four systems shown in Figure 6 (Vela X-1, SMC X-1, Cen X-3, and LMC X-4) are commonly accepted to be accretion–powered pulsars (e.g., X-ray binary systems). Estimates of the masses for these systems are from Nagase (1989). Derived masses of accretion powered systems are more uncertain than for neutron star – neutron star binaries since Keplerian-order observations of both the pulsar's *and* the companion's Doppler velocity curves are required. A measurement of the eclipse duration is also necessary so that the orbital inclination angle can be estimated. Even though X-ray binary neutron stars may constitute a population with a different distribution function than neutron star – neutron star binaries we include them in the analysis. Thorsett et al. (1993) and Brown et al. (1995) give a more complete sample (17 systems) of neutron star mass determinations and show that the average observed neutron star mass is $1.35 \pm 0.27$ $M_\odot$. This summary is plotted and labeled in Figure 6. Eliminating the more uncertain measurements of Her X-1 and Vela X-1, Thorsett et al. (1993) show that the error bars on the neutron star mass narrow considerably to $1.35 \pm 0.1$ M$_\odot$.

The calculated neutron star mass functions are shown in the lower half of Figure 6. The width of the mass bins and normalization procedure is the same as that used for Figure 3. If all massive stars explode as Type II supernovae, the initial remnant mass function is identical to that shown in Figure 3, shown again in Figure 6a as the histogram of filled circles. As noted previously, this mass function has significant peaks with average masses at 1.28 M$_\odot$ and 1.73 M$_\odot$, and an overall average mass of $1.61 \pm 0.21$ M$_\odot$.

Since not all massive stars end their lives as Type II supernova, we consider a simple picture in which the ratio of Type II to Type Ib supernova is a constant independent of initial main–sequence mass and metallicity. Assuming the Milky Way system to be Hubble type Sbc and a Hubble constant of 75 km s$^{-1}$ Mpc$^{-1}$, van den Bergh & McClure (1994) estimate that 1/4 to 1/2 of all core collapse events are Type Ib supernova depending on the luminosity function. We repeated our calculations using the Type Ib supernova models of Woosley et al. (1995) as a guide to the expected remnant masses for stars of various helium core mass and assuming that a certain constant fraction at each stellar mass produces remnants given by Woosley et al. (1995) instead of Woosley & Weaver (1995). The Type Ib models differ from the Type II models chiefly in that the former have a greatly reduced helium core mass. This reduces the mass and binding energy of the mantle, with a corresponding reduction in the amount of material that falls back. Lack of a hydrogen envelope ensures the Type Ib models do not undergo the reverse shock phenomena which also indicates less fallback material.

The helium cores of Woosley et al. (1995) were mapped into a main–sequence mass by assuming, for example, that a 4 M$_\odot$ helium core corresponds to a 15 M$_\odot$ star. Similarly, we took 5 M$_\alpha$ = 18 M$_\odot$, 7 M$_\alpha$ = 22 M$_\odot$, 10 M$_\alpha$ = 26 M$_\odot$, and 20 M$_\alpha$ = 45 M$_\odot$. This transformation assumes that the helium core is uncovered almost at the start of helium burning, which for stars of less than 35 M$_\odot$ generally requires a close binary companion *and* an early transition to a red supergiant. Hydrogen core and shell burning phases in stars of different mass can lead to significant variations in the final mass of the helium core, and our principal results are mildly sensitive to variations in the adopted main–sequence to helium core mapping.

Under the assumptions of this simple model, if 1/2 of all massive stars explode as Type II events and 1/2 as Type Ib supernova, then the peak at 1.27 M$_\odot$ in the Type II neutron star birth function spreads out. This distribution is shown in Figure 6b as the filled triangle histogram. Making the unrealistic assumption that *all* massive stars end up as Type Ib supernova gives a narrow spread of neutron star masses (average mass of 1.33 M$_\odot$ with a standard deviation 0.05 M$_\odot$ centered at 1.27 M$_\odot$) and is shown as the filled square histogram in Figure 6c. Type Ib fractions intermediate to the three cases we have considered follow by interpolation. The reason the low–mass peak in the distributions spread out as the fraction of Type Ib supernova is increased is that the Type Ib progenitors lose all of their hydrogen envelopes and end up with significantly smaller helium cores that cluster around $4 \simeq 6$ $M_\odot$ (e.g., Ensman & Woosley 1988). However, there are only 5 Type Ib progenitors in the Woosley et al. (1995) study, and caution is advisable in interpreting Figures 6b and 6c since the spreads are dominated by small number statistics and the uncertainties of mass loss.

If all stars between 8 and 11 $M_\odot$ become AGB stars, instead of all becoming Type II supernovae that leave a neutron star, then the amplitude of the low–mass peak around 1.27 $M_\odot$ in Figure 6 is drastically reduced and the peak around 1.76 $M_\odot$ becomes the largest (for a classic Salpeter initial mass function). The amplitude and spread around the peaks for cases where a certain fraction of stars between 8 and 11 $M_\odot$ become supernovae lie between the two extremes (all Type II or all AGB) we have considered here.

The principal point of Figure 6 is that all three calculated neutron birth functions are in accord with the observations, with perhaps a slight preference for 1/4 to 1/2 of all massive stars becoming Type Ib instead of Type II supernovae. Although the nuclear equation of state may allow a neutron star masses between 0.2 and 2.5 $M_\odot$, the combined weight of the observational evidence and core collapse supernova calculations suggests that nature prefers to form neutron stars in a narrow range just below 1.4 $M_\odot$. However, all neutron stars with determined masses are, of necessity, in binary systems. This may be why the observed mass range is narrower than what one obtains for a broad range of main–sequence masses evolved as single stars (Brown & Bethe 1994; G. E. Brown 1995, private communication). In addition, with the high birth velocity observed for radio pulsars (Lyne & Lorrimer 1994) binary systems with either a Type II or a Type Ib component may frequently become unbound unless there is some small asymmetry to the explosion (Burrows & Woosley 1986). It may be prudent in future theoretical studies to distinguish between those systems that remain bound from those binaries that become unbound.

The lightest neutron star that could be produced in any of the Type II models was 1.18 $M_\odot$ and 1.22 $M_\odot$ for the Type Ib models (gravitational masses). This is in good agreement with neutron star – neutron star measurement of $1.2 \pm 0.26$ $M_\odot$ for 2303+46 and the lower limit of 1.2 $M_\odot$ for 1713+0747, but only in fair agreement with the values for the X-ray binary systems SMC X-1, 1538-522, Cen X-3, and Her X-1 (1.17, 1.06, 1.09, and 1.04, respectively) although the uncertainty in these measurements is much larger (Thorsett et al. 1993, Brown et al. 1995).

Finn (1994) employed Bayesian statistical techniques on the binary pulsar observations (ignoring the X-ray binary systems) and claimed a 95% confidence upper neutron star mass limit of 1.64 $M_\odot$. Finn assumed that the neutron star masses are uniformly distributed between 0.1 and 3.0 $M_\odot$, in sharp contrast to the distinct bimodal distributions found in our detailed stellar-chemical evolution calculations. However, our peak at 1.76 $M_\odot$, which includes any isolated neutron stars, is not inconsistent with Finn's estimate of the upper neutron star mass limit in neutron star – neutron star binaries.

A number of groups have shown that, under certain circumstances, a slowly accreting white dwarf composed either of carbon and oxygen or of neon and oxygen can collapse directly to a neutron star with little or no accompanying optical emission (Nomoto et al. 1979; Canal, Isern, & Labay 1980; Nomoto 1986; Woosley & Weaver 1986; Baron et al. 1987; Mayle & Wilson 1988; Canal, Isern, & Labay 1990; Nomoto & Kondo 1991; Isern, Canal, & Labay 1991). Accretion–induced collapse has also been invoked to explain certain classes of X-ray binaries (Canal et al. 1990; van den Heuvel 1984, 1987) and the formation of millisecond pulsars (Bailyn & Grindlay 1990). The parameters determining the fate of the accreting white dwarf are the central density and the speed with which the conductive deflagration propagates. Accretion–induced collapses of Chandrasekhar mass mass white dwarfs should yield a unique neutron star mass: 1.39 $M_\odot$ baryonic which becomes a gravitational mass of 1.27 $M_\odot$ after subtracting the binding energy (eq. 8), although this unique mass could be increased if significant accretion continues after formation of the neutron star. A significant population of neutron stars formed from accretion–induced collapse would change the neutron star birth functions shown in Figure 6.

## 6. DEPENDENCE ON THE STELLAR INITIAL MASS FUNCTION

The amplitude of the two peaks centered at 1.27 $M_\odot$ and 1.76 $M_\odot$ in the Type II distribution is shown as a function of the exponent in a Salpeter initial mass function in Figure 7. Mass resolution and normalization procedures are the same as those used in Figure 3. For a relatively flat initial mass function, exponent of -1.0, the peak at 1.76 $M_\odot$ is $\simeq$ 20% larger than the peak at 1.27 $M_\odot$. When the exponent is -1.15 the amplitudes of the two peaks are equal. For a sharply decreasing initial mass function, exponent of -2.0, the peak at 1.27 $M_\odot$ is about twice as large than the peak at 1.76 $M_\odot$. There are simply less of the most massive stars as the exponent becomes more negative. The dashed lines in Figure 7 indicate the classical Salpeter exponent of -1.35 (the value used in all the previous figures) and the effective exponent of -1.9 in the Miller & Scalo (1979) initial mass function. Hence, if we had employed the Miller & Scalo (1979) initial mass function or chosen a steeper slope to the Salpeter initial mass function in the previous figures, then the amplitude of 1.76 $M_\odot$ peak would have been even smaller, possibly becoming statistically insignificant.

On the other hand, if all stars between 8 and 11 $M_\odot$ become AGB stars, instead of all becoming Type II supernovae that leave a neutron star, the situation is quite different.

In this case, for a relatively flat initial mass function, exponent of -1.0, the peak at 1.76 $M_\odot$ is $\simeq 50\%$ larger than the peak at 1.27 $M_\odot$. Only when the exponent is -1.55 are the amplitudes of the two peaks are equal. For a sharply decreasing initial mass function, exponent of -2.0, the peak at 1.27 $M_\odot$ is $\simeq 50\%$ larger than the peak at 1.76 $M_\odot$.

## 7. $\Delta Y / \Delta Z$ AND BLACK HOLE CONSTRAINTS

Often a hypothesis of a linear relationship between the helium content and the metallicity content is invoked

$$Y \; = \; Y_p \; + \; \left(\frac{\Delta Y}{\Delta Z}\right) \; Z \; . \qquad (9)$$

The observed helium and overall metal abundances influence estimates of the primordial helium $Y_P$ originating from Big Bang nucleosynthesis (the intercept) and the amount of nuclear processing done by stars (the slope). The assumption of linear relationship is only valid at low metallicities. At metallicities larger than $\simeq 0.1$ $Z_\odot$, the contributions from Type 1a supernovae and low–mass stars cause some curvature. The best constraints on $\Delta Y/\Delta Z$ come from low–metallicity Galactic and extragalactic H II regions (Peimbert 1986; Pagel, Terlevich, & Melnick 1986; Pagel et al. 1992). Reduction of the spectrographic data is especially demanding in these observations. The principle sources of uncertainty are the removal of contamination from Wolf-Rayet stars, correction for the amount of neutral helium inside the H II regions, problems from telluric and Galactic absorption lines, disagreement on the theoretical recombination coefficients, corrections for the high population of the metastable He I 2 $^3$S state, transformation of the observed O/H or O/N ratios into total metallicity values, and the effects of oxygen depletion through grain formation (Peimbert 1986; Osterbrock 1989; Pagel et al. 1992). Although the variability is large, values of $\Delta Y/\Delta Z = 4.0 \pm 1.0$ are taken to be representative of the observations.

The quantity $\Delta Y/\Delta Z$ is sensitive to the maximum mass of a star that will give its nucleosynthetic products back to the interstellar medium. Massive stars produce some helium, but mostly metals, of which oxygen comprises the vast bulk. The metal-poor exploded massive star models of Woosley & Weaver (1995) have $\Delta Y/\Delta Z \simeq 1.3$, with previous surveys indicating similar values. If all the material in massive stars is returned to the interstellar medium, then it is difficult to satisfy the observed low–metallicity $\Delta Y/\Delta Z$ ratio. As a result, several authors have suggested that the most massive stars become black holes and swallow a good fraction, if not all, of their metals. Integrating over a Salpeter type initial mass function, Maeder (1992,1993) suggested that the observed $\Delta Y/\Delta Z$ is best reproduced if black holes are formed above about $\simeq 20$ $M_\odot$, while Brown & Bethe (1994) suggested $25 \pm 5$ $M_\odot$ using the same formalism. With a detailed stellar–chemical evolution model, Timmes et al. (1995) found that the suggested ratio could be fitted if stars above $\simeq$ 30 M$_\odot$ become black holes. The tentative conclusion is that the Galaxy could contain a large number of stellar mass black holes. However, it may be myopic to examine only the $\Delta Y/\Delta Z$ ratio and ignore the effects of a black hole cutoff on the evolutionary histories of all the elements. This concern, combined with the large scatter present in the observations and the difficult reduction of the spectroscopic data, suggests that strong statements regarding the black hole formation mass that are strictly based on $\Delta Y/\Delta Z$ should be viewed with caution.

## 8. CONCLUSIONS

The number of neutron stars as a function of their mass has been determined from realistic calculations of supernova models and Galactic chemical evolution. We find that Type II supernovae give rise to a bimodal distribution of initial neutron star masses. The distribution has two peaks at 1.27 M$_\odot$ and 1.76 M$_\odot$ (gravitational masses) and these two peaks have average masses of 1.28 $\pm$ 0.06 and 1.73 $\pm$ 0.08 M$_\odot$. For a Salpeter (x = -1.35) initial mass function, the peak at 1.27 M$_\odot$ has the largest amplitude, but this result is sensitive to the behavior of stars in the narrow 8 $\sim$ 11 M$_\odot$ range and to the Type Ib supernova. These stars control the amplitude, and to a lesser extent the width, of the remnant peak clustered around 1.27 M$_\odot$. If the vast majority of stars in this mass range become AGB stars, then the peak centered at 1.76 M$_\odot$ has the largest amplitude. Type Ib supernovae chiefly cause the lighter mass peak to spread by a small amount and reduce the relative amplitude of the 1.76 M$_\odot$ remnant peak. The chief reason for the dichotomy in the neutron star masses is a bifurcation in stellar evolution that occurs following helium–burning in stars that do or do not burn carbon convectively in their centers. The critical main–sequence mass for this split is 19 M$_\odot$, a value which depends upon the adopted rate for $^{12}$C$(\alpha,\gamma)^{16}$O. If all of the more massive stars (the peak at 1.76 M$_\odot$) became black holes, then the average neutron star mass for a mixture of Type II supernovae and half as many Type Ib's would be 1.30 $\pm$ 0.06 M$_\odot$.

All these masses are probably lower bounds because, although they do include an uncertain amount of mass that falls back at late times owing to hydrodynamic interaction of the shock with the mantle and envelope of the star, they do not include the mass that accretes onto the iron core as the explosion itself is developing (i.e., the first few seconds) or any mass that subsequently accretes in a binary system. Determining the systematics of the mass that accretes during the explosion will be a difficult task for the future. Based upon current and past studies of the explosion, we expect that it is larger for heavier stars and ranges from 0.01 to $\sim$ 0.2 M$_\odot$. We have also provided (in §3 and Fig. 3) lower and upper bounds based upon the assumption that the final mass separation occurs at either the iron core or oxygen shell.

Our calculated birth function, especially the low–mass peak, is found to be consistent with mass measured for neutron stars in binary systems, notably the average mass of 1.35 ± 0.27 M$_\odot$ determined for 17 systems (see Figs. 3 and 6). Sources of uncertainty were discussed, with the slope of the initial mass function and the dependence of the remnant mass on explosion energy being two of the most important. Variations in the exponent of a Salpeter initial mass function were shown not to affect the locations of the peaks in the neutron star distribution function, but do affect the relative amplitudes of the two peaks. Estimates of the total number of neutron stars due to core collapse events in the Galaxy range from 0.4 to 1.9 × 10$^9$, but depend upon many uncertainties. For reasonable assumptions about the maximum mass of a neutron star and explosion energy, the total number of black holes in the Galaxy is estimated be $\lesssim$ 1.4 × 10$^9$. There should be no neutron stars produced from massive stars lighter than 1.18 M$_\odot$.

This work has been supported at Santa Cruz by the NSF (AST 91 15367 and 94 17171) and NASA (NAGW 2525); at Livermore by the Department of Energy (W-7405-ENG-48); and at Chicago by an Enrico Fermi Postdoctoral Fellowship (F.X.T.).

**Figures and Captions**

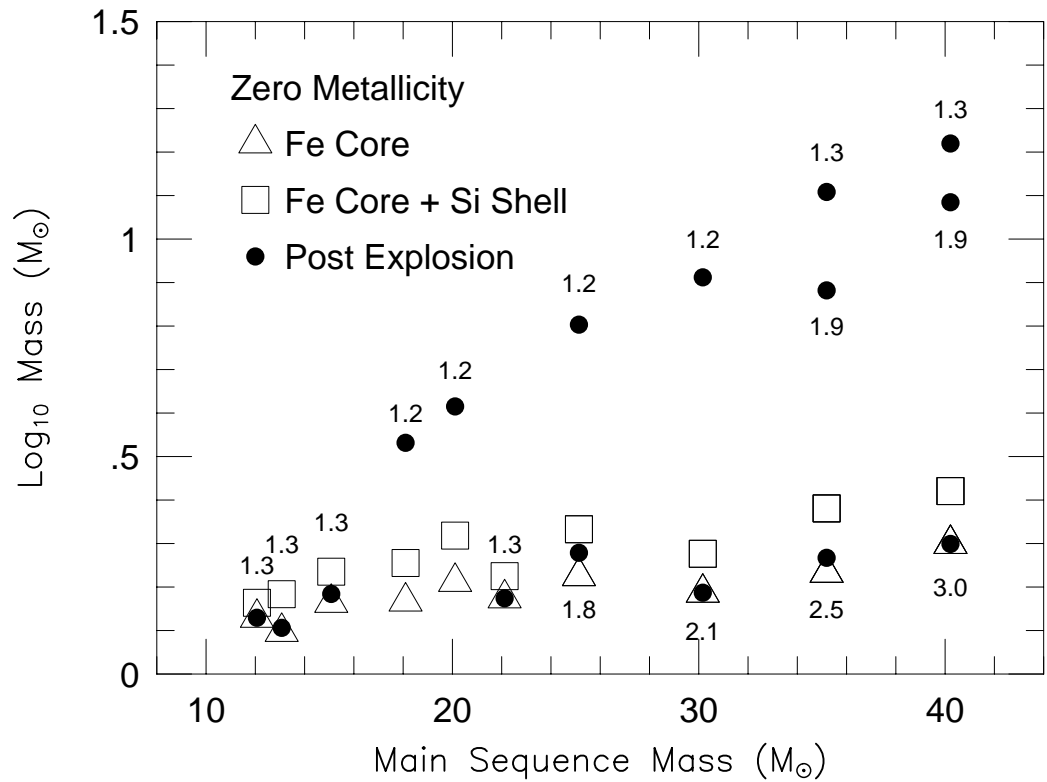

Fig. 1.— Core and remnant masses for zero–metal stars. The iron core masses of the presupernova models are shown as the open triangles, iron core mass plus silicon shell mass (total mass interior to the oxygen–burning shell) of the presupernova stars as the open squares and the final remnant mass of the exploded stars (which includes any fallback material) as the filled circles. Alongside each remnant mass the kinetic energy of the ejecta at infinity is given in units of $10^{51}$ ergs. For $M \geq 25$ $M_\odot$, several explosion energies, hence several remnant masses, were investigated.

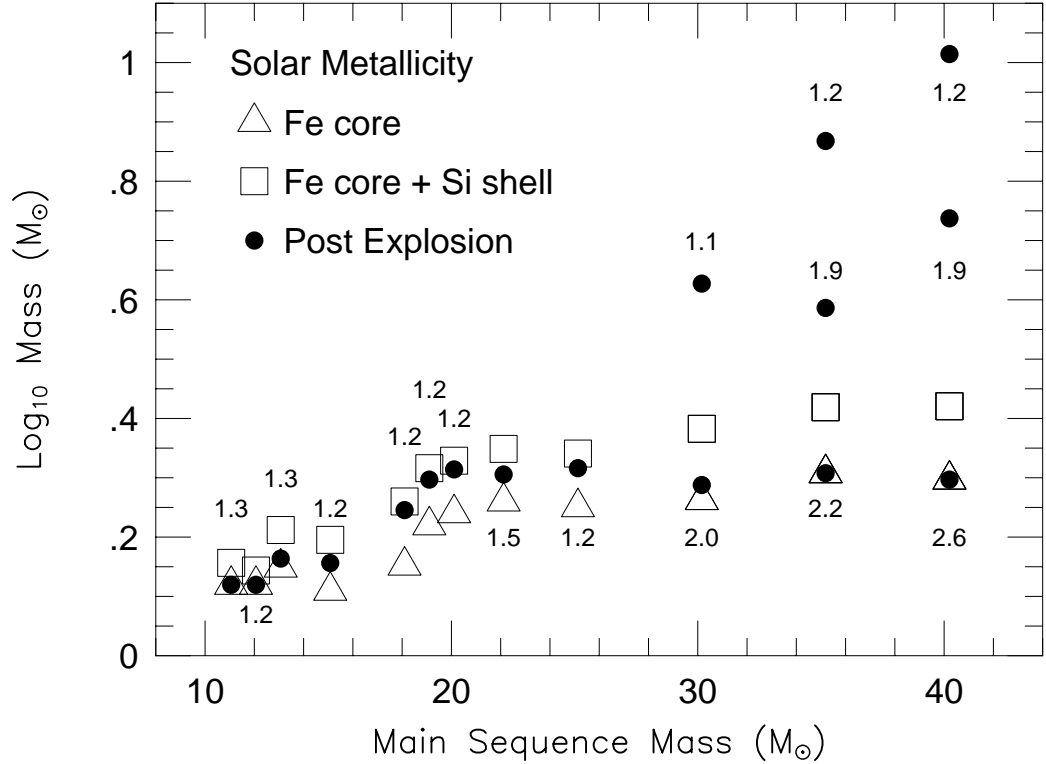

Fig. 2.— Core and remnant masses for solar metallicity stars. The symbol legend and labeling scheme is the same as in Fig. 1. To have roughly equal remnant masses for all main-sequence masses the explosion energy in the M ≥ 30 M$_\odot$ presupernova models must be increased to overcome the binding energy of the mantle. A constant kinetic energy of the ejecta at infinity means increasing amounts of matter falls back onto the remnant. The remnant masses left by solar metallicity stars tend to be smaller than their zero–metal counterparts since the former have a more extended presupernova density stratification.

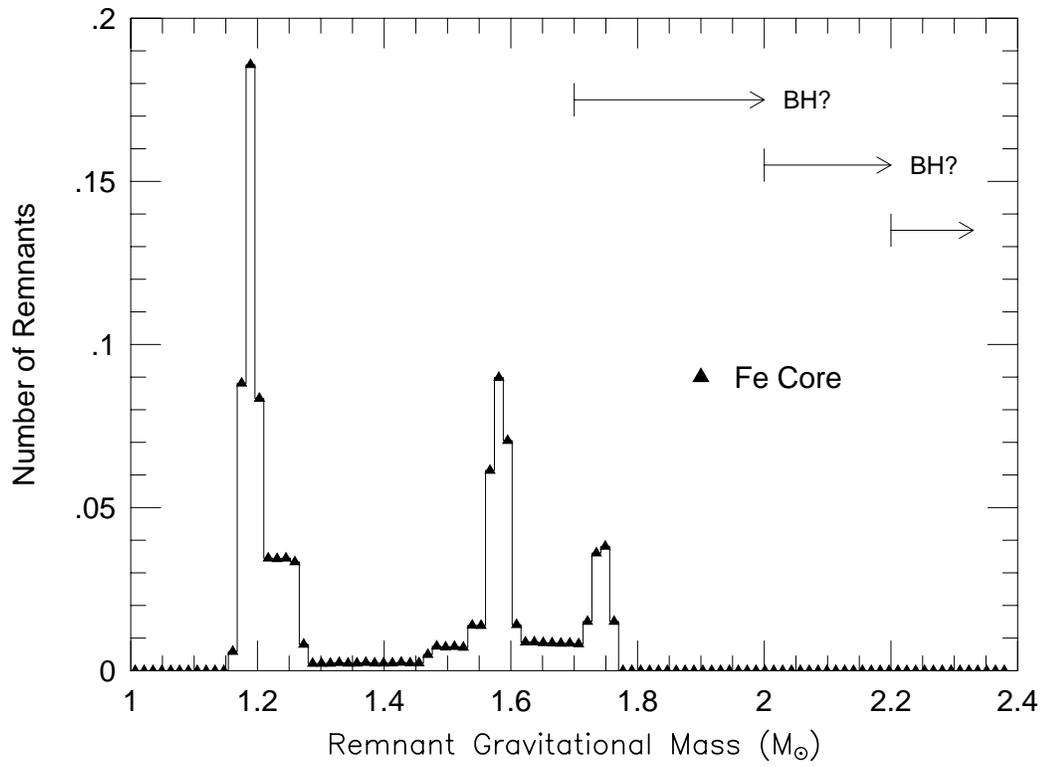

Fig. 3a.

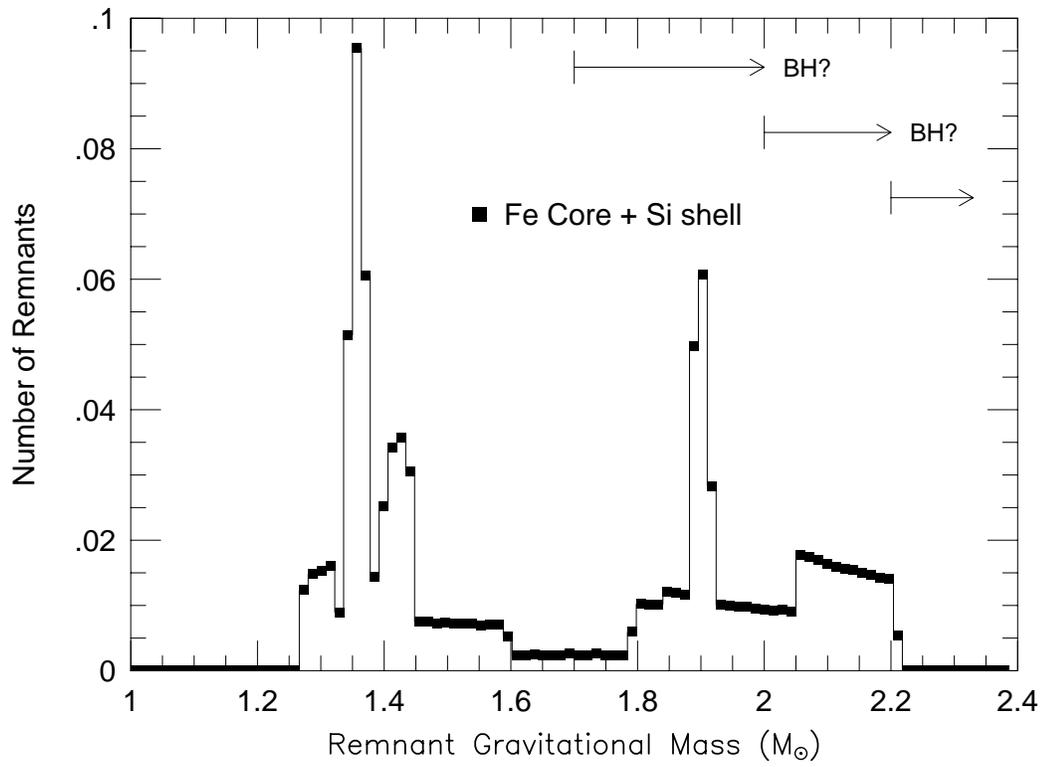

Fig. 3b.

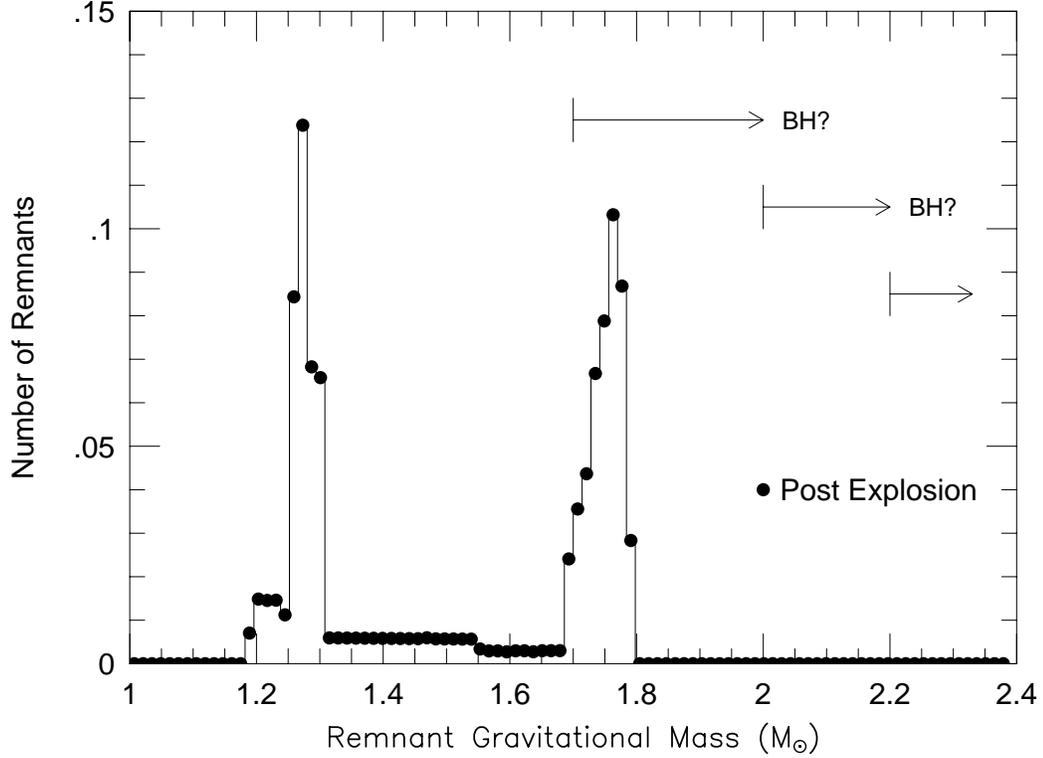

Fig. 3c.
Fig. 3. Remnant distribution functions. The normalized number of remnants as a function of gravitational mass is shown in Fig. 3a assuming that only the presupernova iron core stays behind. This distribution is bimodal, with significant peaks at 1.23 and 1.60 $M_\odot$. Fig. 3b shows a similar plot assuming that the remnant is the iron core plus original silicon shell and has significant peaks at 1.43 and 1.91 $M_\odot$. The largest explosion energy models (see Figs. 1 and 2) show a distribution in Fig. 3c with peaks centered at 1.27 and 1.76 $M_\odot$. The post–explosion masses are between the (lower bound) iron cores and the (rough upper bound) mass interior to the oxygen–burning shell. The relative amplitude, but not the location, of the peaks vary with the choice of the exponent in the initial mass function. If every remnant above $\simeq 1.7$ $M_\odot$ turns into a black hole, then only the peak around 1.27 $M_\odot$ would remain.

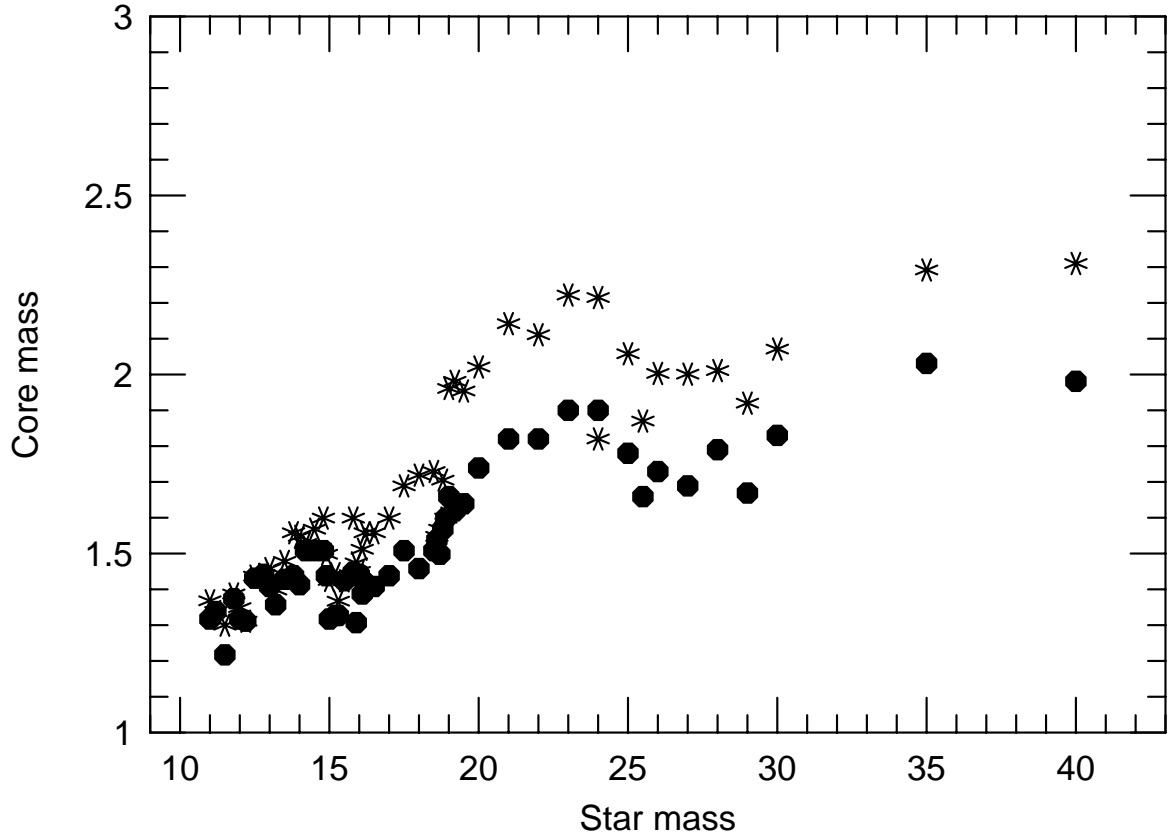

Fig. 4. Iron core masses (filled circles) and iron core masses plus the mass interior to the oxygen–burning shell (open stars) for a fine mass grid of presupernova stars. Superposed upon the overall tendency for a monotonic increase of the iron core mass is an abrupt jump around 19 M$_\odot$. This mass marks the transition between stars that burn carbon convectively in the core and those that burn radiatively (see text).

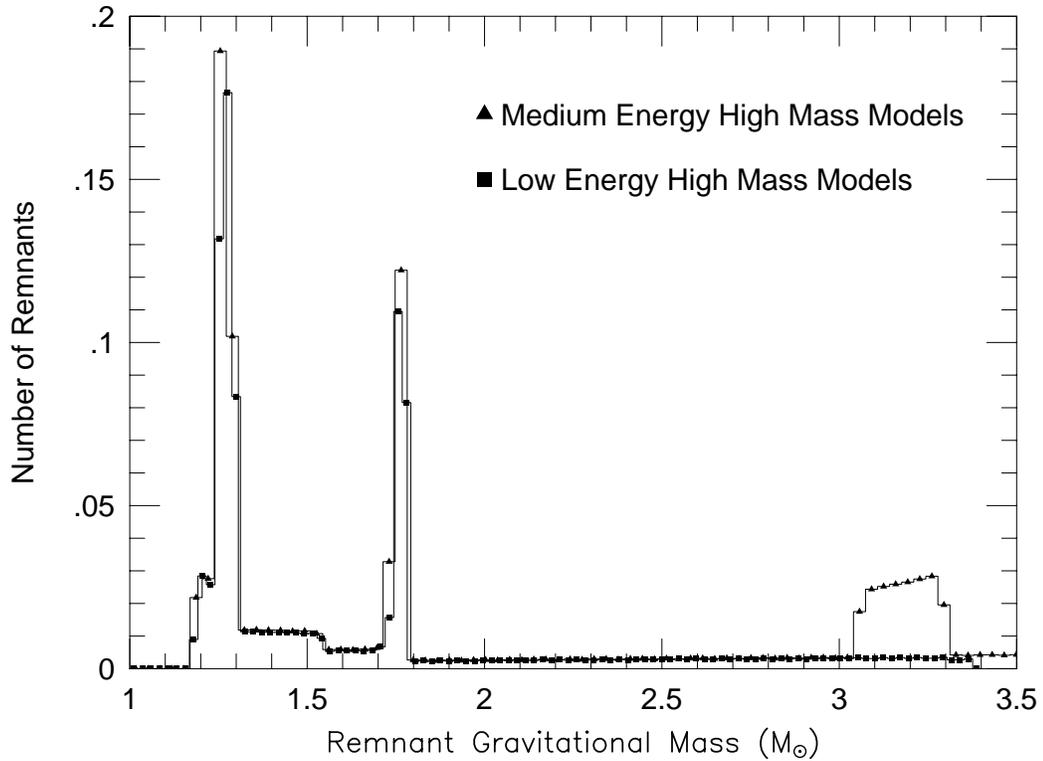

Fig. 5.— Energy dependence of the post–explosion birth functions. The width of the mass bins is the same as in Fig. 3 and the same normalization procedure used. The medium and low–explosion energy models have remnant peaks in roughly the same mass bins as in the high–energy models of Fig. 3. An excess of massive remnants appears near at 3.3 $M_\odot$ for the medium energy explosions. A similar excess exists for the low–energy explosions but spans a large mass range and has a small amplitude (only 1 or 2 in each mass bin). These small excesses at large remnant masses are probably devoid of statistical significance.

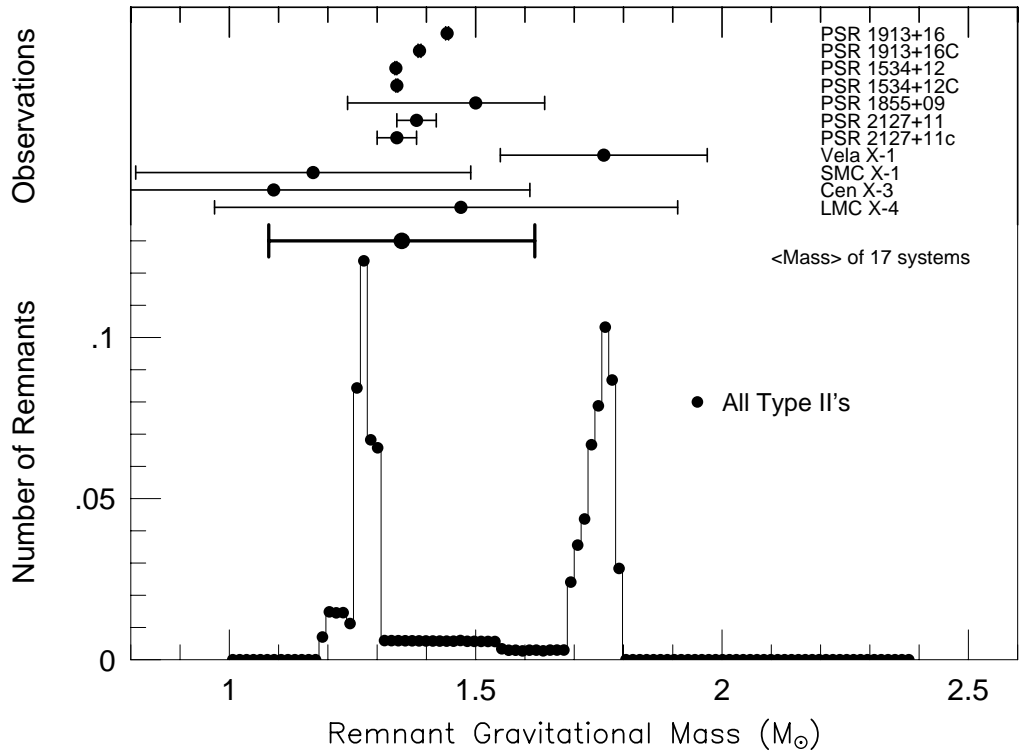

Fig. 6a.

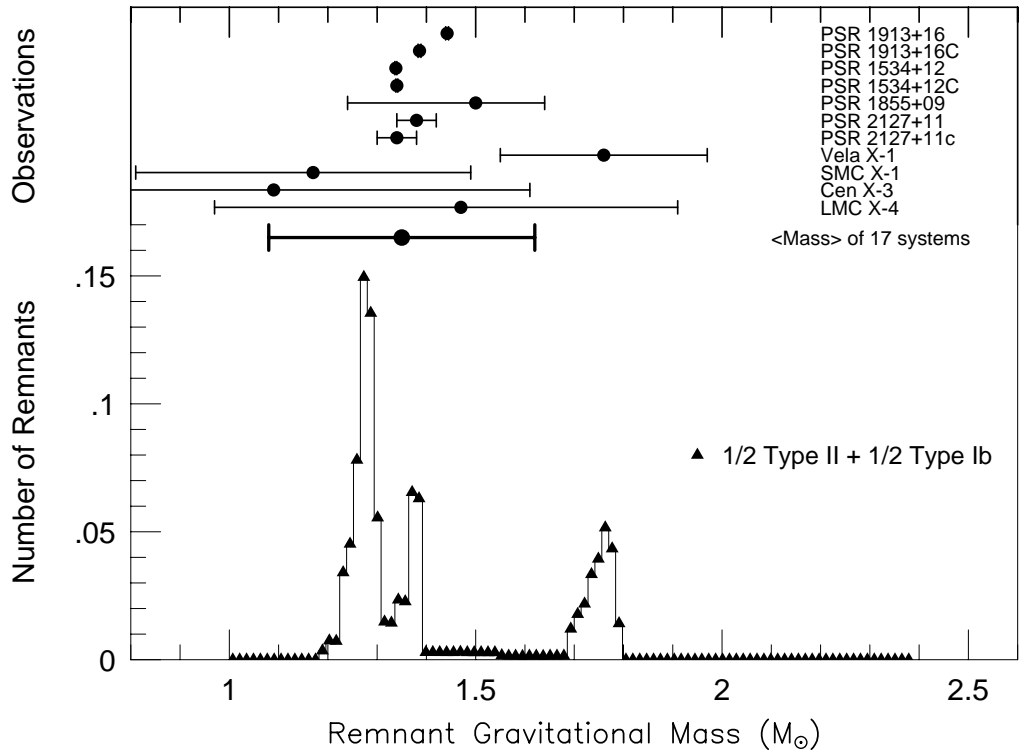

Fig. 6b.

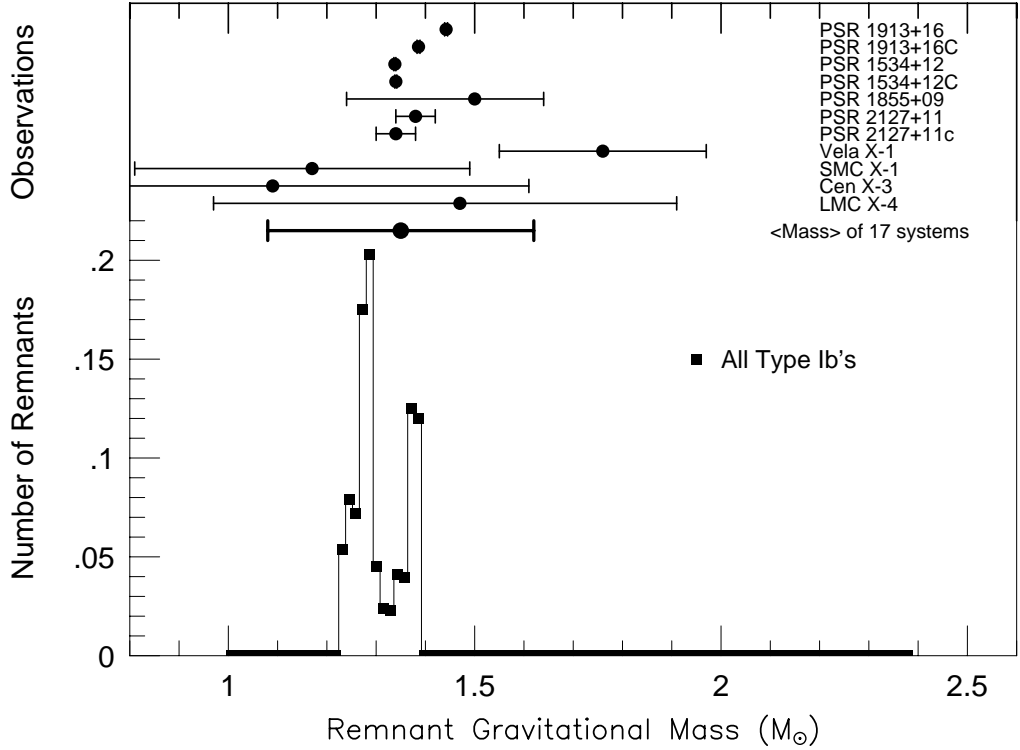

Fig. 6c.

Fig. 6. Measured neutron masses and the theoretical birth functions. Mass determinations in neutron star–neutron star binary systems and X-ray binary systems are shown in the upper half of the figures. Thorsett et al. (1993) and Brown et al. (1995) give a more complete sample than shown here and calculate that the average neutron mass in 17 systems is 1.35 ± 0.27 $M_\odot$. The calculated neutron star birth functions are shown in the lower half of the figures. The case where all massive stars explode as Type II supernovae is shown as the filled circle histogram in Fig. 6a. If one–half of all massive stars explode as Type II events and one–half as Type Ib supernova, then the calculated neutron star birth function is given by the filled triangle histogram (see text) in Fig. 6b. The other extreme where all massive stars end up as Type Ib supernova is shown at the filled square histogram in Fig. 6c. Addition of Type Ib supernovae causes the 1.27 $M_\odot$ peaks of the Type II neutron star birth function to spread out.

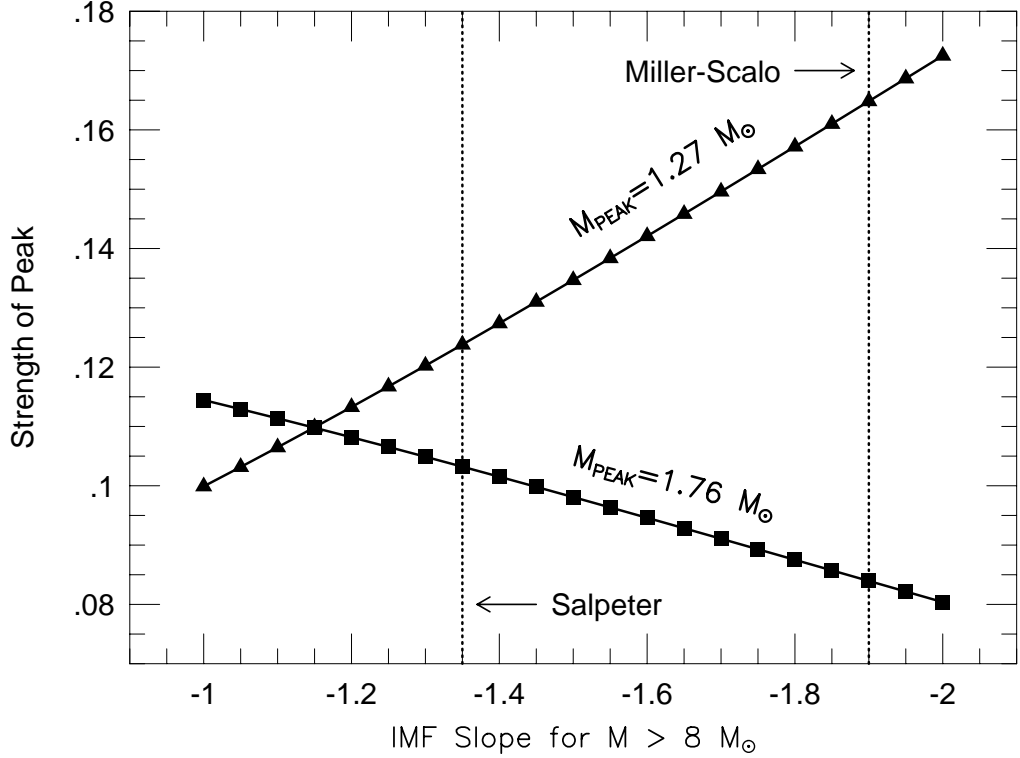

Fig. 7.— Sensitivity of peak amplitudes to the slope of the initial mass function. Widths of the mass bin widths and normalization procedures are the same as those used in Figure 3. For a relatively flat initial mass function, exponent of -1.0, the peak at 1.76 $M_\odot$ is $\simeq 20\%$ larger than the peak at 1.27 $M_\odot$ in the pure Type II distribution. When the exponent is -1.15 the amplitudes of the two peaks are equal. For a sharply decreasing initial mass function with an exponent of -2.0 the peak at 1.27 $M_\odot$ is twice as large large as the peak at 1.76 $M_\odot$. If the Miller & Scalo (1979) initial mass function had been used or a steeper Salpeter initial mass function had been used in the previous figures instead of the classical -1.35 exponent, then the amplitude of 1.76 $M_\odot$ peak would have been even smaller, possibly becoming statistically insignificant.